\documentclass{jpp}
\usepackage{graphicx}
\usepackage{epstopdf, epsfig}
\usepackage{hyperref}

\usepackage[normalem]{ulem}
\usepackage{xcolor}

\shorttitle{Multi-scale magnetic field structures in an expanding elongated plasma\dots}
\shortauthor{M.~A.~Garasev et~al.}

\title{Multi-scale magnetic field structures in~an~expanding elongated plasma cloud with hot electrons subject to an external magnetic field}

\author{M.~A.~Garasev\aff{1},
  A.~A.~Nechaev\aff{1},
  A.~N.~Stepanov\aff{1},
  V.~V.~Kocharovsky\aff{2}
 \and Vl.~V.~Kocharovsky\aff{1}\corresp{\email{kochar@ipfran.ru}}}

\affiliation{\aff{1}Institute of Applied Physics, Russian Academy of Sciences, Nizhny Novgorod, 603950, Russia
\aff{2}Department of Physics and Astronomy, Texas A\&M University, 
College Station, TX 77843, USA}

\begin{document}

\maketitle

\begin{abstract}
We carry out 3D and 2D PIC-simulations of the expansion of a magnetized plasma that initially uniformly fills a half-space and contains a semi-cylindrical region of heated electrons elongated along the surface of the plasma boundary. This geometry is related, for instance, to ablation of a plane target by a femtosecond laser beam under quasi-cylindrical focusing. We find that a decay of the inhomogeneous plasma--vacuum discontinuity is strongly affected by an external magnetic field parallel to its boundary. 

We observe various transient phenomena, including an anisotropic scattering of electrons and an accompanying Weibel instability, and reveal various spatial structures of the arising magnetic field and current, including multiple flying apart filaments of a z-pinch type and slowly evolving current sheets with different orientations. The magnitude of the self-generated magnetic field can be of the order of or significantly exceed that of the external one. Such phenomena are expected in the laser and cosmic plasmas, including the explosive processes in the planetary magnetospheres and stellar coronal arches.
\end{abstract}

\section{Introduction}
\label{sec1}

The problem of generation and coexistence of large- and small-scale magnetic fields during the expansion of a non-equilibrium collisionless plasma with hot electrons into a cold background plasma or vacuum is typical for many situations in the laboratory and cosmic plasma physics, including the phenomena of laser ablation, dynamics of the planetary magnetosheath regions, evolution of the solar (stellar) flares and wind formation \citep[see, e.g.,][]{Huang2019, Srivastava2019, Quinn2012, Romagnani2008, Gode2017, Gruzinov2001, Lyubarsky2006, Garasev2016, Medvedev1999, Spitkovsky2008, Huntington2015, Sakawa2016, Ruyer2020, Balogh2018}. A plasma discontinuity or a transition layer could be inhomogeneously heated and exposed to an external magnetic field comparable in magnitude to the self-generated one and oriented at various angles relative to the direction of the plasma density gradient or the axis of a plasma anisotropy induced by the expansion of a hot electron cloud.

We consider a particular, though quite typical case of a plasma uniformly filling a half-space with a semi-cylindrical region of heated (Maxwellian) electrons elongated along the surface of the plasma boundary. This case is relevant, e.g., to the expansion of a laser plasma produced by a femtosecond laser beam under quasi-cylindrical focusing on a plane target or the injection of a current filament with hot electrons into the upper part of a stellar coronal arch. By means of the 3D and 2D PIC-simulations, we describe the inhomogeneous plasma flow and the evolution of the current structures of different scales, paying particular attention to the anisotropic cooling of electrons and the interplay between the self-generated and external magnetic fields in the dynamics of particles. 

We resolve the formation of the stratified plasma flows and current sheets or filaments with dimensions of the order of or less than the particle gyroradius, i.e., the structures of a kinetic origin and reduced dimensionality, which are responsible for a small-scale stratification of the plasma. These structures and an accompanying reconnection of the magnetic field lines cannot be described within the MHD approximation \citep[cf.][]{Patel2021, Srivastava2019, Priest2014, Plechaty2013, Moritaka2016}. In the laser and cosmic plasmas with weak, rare collisions of particles, the leading mechanism of the formation of relatively small-scale current structures is the Weibel instability \citep{Weibel1959, Davidson1989, Kocharovsky2016, Borodachev2017}. It is caused by the anisotropy of the charged particle velocity distribution and can initiate or change various transient processes involving the generation of strong small-scale magnetic fields.

In this work we continue our recent studies \citep{Nechaev2020,Nechaev2020FizPlasm} of the Weibel instability during the decay of a strong discontinuity in the density and temperature of a non-relativistic plasma with hot electrons by means of numerical simulations using the code EPOCH \citep{Arber2015}. Investigating the plasma expansion into vacuum (in this case the electrostatic shock wave does not form), we focus mainly on the role of a uniform external magnetic field that is parallel to the plane boundary of the plasma and oriented either along or across the inhomogeneously heated region elongated along this boundary. To our best knowledge, the decay of a magnetized plasma--vacuum discontinuity in this geometry has not been studied in detail \citep[cf.][]{Moreno2020, Dieckmann2018, Fox2018, Thaury2010, Schoeffler2018}, although under other conditions the Weibel-type instabilities and, in particular, the filamentation instability have been frequently investigated numerically, e.g., for colliding plasma flows, including magnetized ones \citep[see][]{Spitkovsky2008, Chang2008, Sironi2013, Sironi2009, Bret2009, Silva2006, Dieckmann2009, Ruyer2015}.

The strongly non-equilibrium expansion of a magnetized plasma from an initially heated region is possible only if the energy density of an external magnetic field $B_0$ is less than or of the order of the kinetic energy density of the hot electrons. Let $n_0$ and $T$ denote the initial number density and isotropic temperature (in energy units) of the hot electrons, respectively. In the limit $B_0^2 / 8\pi \ll n_0 T$, the field weakly affects the profile of the expanding plasma density, but can significantly influence the arising anisotropy of the electron velocity distribution and the accompanying Weibel-type instability. In the considered geometry, the latter can develop even in the absence of an external field.

The contents of the paper is as follows. Section~\ref{sec2} describes the initial geometry of the plasma--vacuum discontinuity in our PIC-simulations of its decay in the presence of an external magnetic field of various orientations. A typical example of the results of the 3D3V simulations of this plasma discontinuity decay is presented in section~\ref{sec3}. Section~\ref{sec4} describes the main features of such a decay which are clarified by means of 2D3V simulations for a representative set of the external magnetic fields and various plasma densities. In section~\ref{sec5} we discuss the revealed qualitative patterns and some open problems of the phenomenon in question. General conclusions are stated in section~\ref{sec6}.

\section{An initial-value problem of the plasma--vacuum discontinuity decay in the case of a semi-cylindrical region of heated electrons}
\label{sec2}

\begin{figure}
	\centerline{\includegraphics[width=0.7\textwidth]{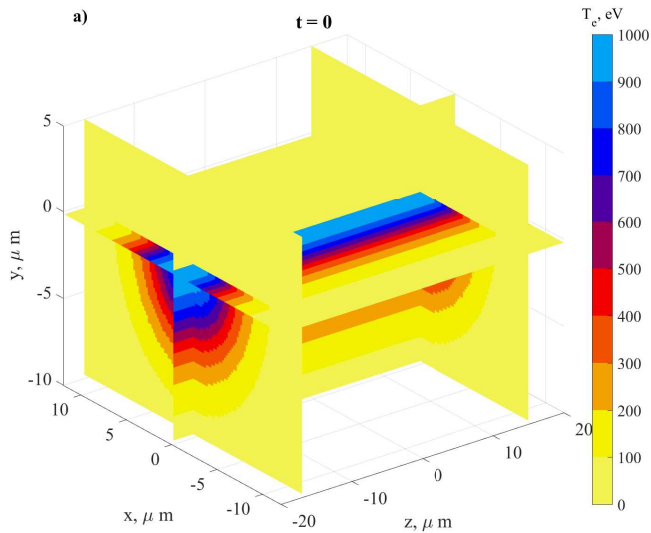}}
    \caption{Geometry of a semi-cylindrical cloud (below the target surface $y = 0$) of heated electrons and their initial temperature distribution in the PIC-simulations. 
    }
\label{fig1}
\end{figure}

Bearing in mind a typical laser-ablation experiment, at the initial moment of time we choose the following density and temperature of ions (with a mass of $M = 100\,m$, where $m$ is the electron mass), placed below the plane $y = 0$ in Cartesian coordinates: $n(y > 0) = 0$, $n(y \leq 0) = $ $n_0 = 1.7 \cdot 10^{22}$, $10^{21}$, or $10^{20}$~cm$^{-3}$ (in different simulations) and $T_\mathrm{i} = 10$~eV. We take the initial temperature of heated Maxwellian electrons (see fig.~\ref{fig1}) to be independent of the coordinate $z$ and vary with the radius according to the Gaussian law with the maximum temperature $T = 1$~keV at the point $x = 0$, $y = 0$ and the asymptotic value $T_{\mathrm{e},\infty} = T_\mathrm{i} = 10$~eV at a large distance $r = (x^2+y^2)^{1/2}$ from that point: $T_\mathrm{e} = T_{\mathrm{e},\infty} + (T - T_{\mathrm{e},\infty}) \exp\left(-r^2/r_0^2\right)$. We assume that $r_0 = 25$~$\mu$m in the simulations with $n_0 = 10^{20}$, $10^{21}$~cm$^{-3}$ and $r_0 = 5$~$\mu$m in the simulations with $n_0 = 1.7 \cdot 10^{22}$~cm$^{-3}$. The geometry of this temperature distribution corresponds to a long semi-cylinder that has the axial section lying on the surface of the plasma, $y = 0$, and the axis directed along the $z$-axis. The external magnetic field $\vec{B}_0$ is directed along either the $y$- or $z$-axis.

Below we describe the results of typical simulations and the revealed physical phenomena for the initial density of heated electrons $n_0$ (equal to the density of ions) within the $10^{21}$ -- $1.7 \cdot 10^{22}$~cm$^{-3}$ range for strong external magnetic fields in the range of 13--2500~T and a lower density, $n_0 = 10^{20}$~cm$^{-3}$, for milder fields in the range of 0.5--13~T. A typical field value of 13~T is chosen based on the parameters of the experiment planned in the Institute of Applied Physics of the Russian Academy of Sciences where an observation of the multi-scale magnetic field structures is possible. For the indicated densities, according to estimates (see also section~\ref{sec4}), the external fields less than 1~T and 0.1~T, respectively, have little effect on the energy and structure of the magnetic fields generated in the plasma. The external fields above 2500~T and 200~T, respectively, practically exclude the explosive decay of a hot plasma discontinuity with the parameters mentioned above.

PIC-simulations using the EPOCH code \citep{Arber2015} are carried out either as completely three-dimensional 3D3V with periodic boundary conditions at $z = \pm L_z / 2$ (with $L_z = 40$~$\mu$m) or simplified 2D3V, omitting all dependencies on the coordinate $z$, but still taking into account all three components of all vectors (including the particle velocity vectors). In the latter case, the $z$-axis is directed across the computational plane $xy$ and the development of the Weibel instability or the plasma-boundary currents, as a rule, leads to the formation of filaments or current sheets elongated mainly along this $z$-axis. On the side boundaries of the computational domain ($x = \pm L_x / 2$), parallel to the $y$-axis, the periodic boundary conditions for particles and fields are used. The lower boundary ($y = - L_y / 4$) reflects particles but allows the fields to escape (be absorbed). The upper boundary ($y = 3 L_y / 4$) is open for both particles and fields. The dimensions are $L_x \times L_y = 240 \times 240$~$\mu$m$^2$ for the density $n_0 = 10^{20}$, $10^{21}$~cm$^{-3}$ and $L_x \times L_y = 36 \times 36$~$\mu$m$^2$ for the density $n_0 = 1.7 \cdot 10^{22}$~cm$^{-3}$.

The plasma comprises $2.5 \cdot 10^9$ ($2 \cdot 10^8$) macroparticles of each fraction, electrons and ions, in 3D3V (2D3V) calculations and the computational domain consists of $400 \times 400 \times 400$ ($1200 \times 1200$) cells. The simulation duration is mainly limited by the moment of time $\tau_\mathrm{R} = 6 \cdot 10^4 \, \omega_\mathrm{pe}^{-1}$ when the transient phenomena have been already fully manifested but the qualitative differences between the 3D3V and simplified 2D3V calculations have not yet usually arisen. Here $\omega_\mathrm{pe} = (4 \pi e^2 n_ 0 / m)^{1/2}$ is the plasma frequency and $e$ is the elementary charge. We have $\tau_\mathrm{R} \approx$ 8~ps and 100~ps for $n_0 = 1.7 \cdot 10^{22}$ and $10^{20}$~cm$^{-3}$, respectively. The skin depths are $c / \omega_\mathrm{pe} \approx$ 0.04~$\mu$m and 0.5~$\mu$m, respectively, and do not appear in our simulations, so that the observed phenomena are not related to this scale.

\begin{figure}
	\centerline{\includegraphics[width=1.0\textwidth]{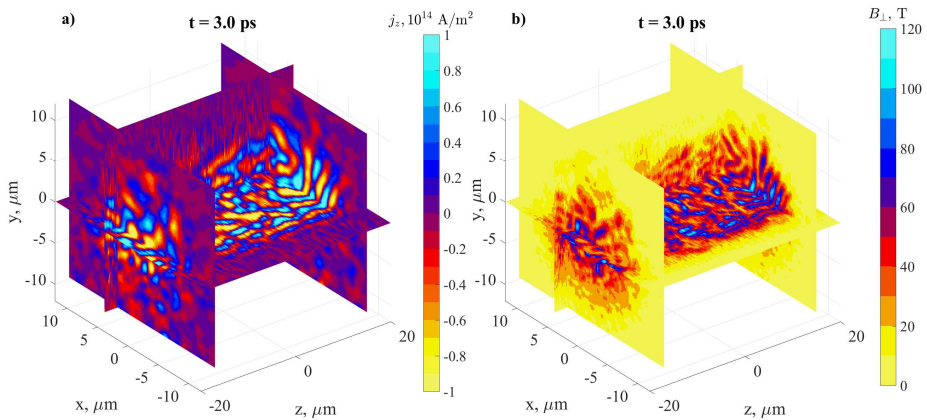}}
    \caption{
    (\textit{a}) The current density component $j_z$ representing the self-generated filamentary structure that is mostly aligned with the $z$-axis and (\textit{b}) the magnitude of the transverse magnetic field, $\left( B_x^2 + B_y^2 \right)^{1/2}$, from the 3D PIC-simulation of the expansion of the plasma cloud with hot electrons into vacuum at the moment of time $t = 3.0$~ps $\approx 3 \cdot 10^4 \omega_\mathrm{pe}^{-1}$. The initial plasma density is $n_0 = 1.7 \cdot 10^{22}$~cm$^{-3}$. The external magnetic field is oriented along the $z$-axis, $\vec{B}_0 \| Oz$, $B_{0z} = 250$~T.
    }
\label{fig1b}
\end{figure}

For the chosen parameters, particle collisions are weak in the rarefied plasma regions of interest where the expansion into the vacuum (towards $y > 0$) and the formation of currents take place. Indeed, already for the densities of $0.1 \, n_0 = 10^{19}, 1.7 \cdot 10^{21}$~cm$^{-3}$ the mean free path of hot electrons, $L_\mathrm{f}$, is within the range 5000--50~$\mu$m, thus greater than $r_0$, $L_{x,y}$ and a typical size of the region where the small-scale magnetic fields are formed (see, e.g., figs.~\ref{fig1b}--\ref{fig6}). Particle collisions are significant in the dense plasma below the target surface where the collisionless PIC-simulations may not be valid.

\section{3D simulation for a typical set of plasma parameters with the~external magnetic field parallel to the elongated heated region }
\label{sec3}

Due to a lack of room, we illustrate a full 3D calculation by just one fig.~\ref{fig1b}. Fig.~\ref{fig1b}b shows a typical tufted structure of the transverse component of the self-generated magnetic field $B_{\bot} = (B_x^2 + B_y^2)^{1/2}$ at a high initial density $n_0 = 1.7~\cdot~10^{22}$~cm$^{-3}$ of hot electrons with an average temperature $T \sim 0.5$~keV and a strong external magnetic field $B_{0z} = 250$~T directed along the axis $z$ of the heated semi-cylinder. 
This distribution of the field is created by the electric current filaments, on average parallel to~$z$  and shown in fig.~\ref{fig1b}a.
At the short time $t = 3.0$~ps after the expansion begins, there is only a small semi-cylinder with a radius less than 3~$\mu$m where the longitudinal self-generated field $B_z - B_{0z}$ is of the order of (and directed opposite to) the external field. At this time and during further expansion at the picosecond timescale, when the plasma pressure dominates over the magnetic one, the displacement of the external field by the large-scale transverse electron currents (as in a solenoid) occurs approximately with an ion-acoustic speed $\propto (T / M)^{1/2} \sim 10^{6}$~m/s, i.e., the deceleration of the plasma flow in the $xy$-plane only slightly exceeds its deceleration in the absence of the external field. (In actual experiments with heavier ions of $M \sim 50000 \, m$ and higher electron temperatures of $T \sim 5$~keV, the plasma will expand with a smaller ion-acoustic speed $\sim 10^{5}$~m/s.) 

Hereafter attention should be paid first of all to the rather universal effect of a small-scale structuring, in particular pinching, of electric currents and the magnetic fields generated by them due to the Weibel-type instability, which itself results from the growth of the electron velocity distribution anisotropy in the nonequilibrium plasma expanding into vacuum. Namely, the temperature of electrons along the $z$-axis practically does not change for a long time, while their temperature in the transverse $xy$-plane decreases rather rapidly and significantly (this is a 3D version of the mechanism originally proposed in \citet{Thaury2010} and further studied in \citet{Nechaev2020}). The instability has the maximum growth rate for the perturbation wave vectors orthogonal to the direction of the maximum electron temperature (the $z$-axis here), and the growth rate decreases \citep{Vagin2014} with a decrease of the angle between a wave vector and the $z$-axis. So, at later times when the growth of the $z$-oriented filaments is saturated due to the self-generated magnetic field, the oblique filaments become quite pronounced and the entire current structure becomes more irregular and strongly dependent on the $z$-coordinate. It is shown already in fig.~\ref{fig1b}a where there are some oblique filaments of finite lengths, especially in the center region of the expanding plasma. 

We justify this conclusion on the Weibel-type mechanism of the growth and saturation of the small-scale current filaments by comparing the gyroradius and gyrofrequency of hot electrons in the self-generated magnetic field with the scale of inhomogeneity and the growth rate of the magnetic field characteristic of the Weibel instability \citep{Vagin2014, Kocharovsky2016, Nechaev2020}. 
As the simulations show, in the region of interest, $0 < y < 5$~$\mu$m (see fig.~\ref{fig1b}), the plasma density and the longitudinal electron temperature are $n_\mathrm{e} \sim 0.1 n_0 \approx 1.7 \cdot 10^{21}$~cm$^{-3}$ and $T_z \sim 0.5$~keV, respectively. Adopting also the observed values of the self-generated transverse field $\sim 50$~T and the temperature anisotropy $A = T_z / T_x - 1 \sim 1$, we find that the gyroradius of a hot electron, $r_B = (2 T_z / m)^{1/2} \omega_{B}^{-1} \approx 1.5$~$\mu$m, is approximately equal to the optimal scale of the electron Weibel instability, $\lambda \sim 10 \, c / \omega_\mathrm{pe} \, (n_0 / n_\mathrm{e})^{1/2} A^{-1/2} \sim 1.4$~$\mu$m, and its gyrofrequency $\omega_{B} \approx 9 \cdot 10^{12}$~rad/s is close to the maximum Weibel growth rate $\Gamma \sim 3 \, (T_x / m)^{1/2} \lambda^{-1} (1 + A^{-1})^{-1} \sim \omega_\mathrm{pe} / 900 \sim 8 \cdot 10^{12}$~s$^{-1}$.
Hence, the magnetic field value satisfies the well-known saturation condition for the Weibel instability \citep[see, e.g.,][]{Kocharovsky2016}. Its characteristic growth time $\tau_B$ in simulations is also consistent with the Weibel mechanism: $\tau_B \sim 10\,\Gamma^{-1}$. 
According to the next section, similar qualitative conclusions ($\lambda \sim r_B$ and $\tau_B \sim 10\Gamma^{-1}$) follow from the 2D simulations (fig.~\ref{fig2}), which have been compared with the 3D simulations and shown to be correct with respect to the main features of the small- and large-scale structures of the generated magnetic fields.

Also noticeable is the cumulation (focusing) of the plasma flow, formed collectively by the hot electrons and the cold ions, in the direction of the maximum deformation of the magnetic field lines in the region of the greatest pressure of escaping electrons. The cumulation takes place when this field is oriented along the long heated region on the target surface as is discussed in the next section (see fig.~\ref{fig2}). Note that in all of these simulations the dynamics of electric currents is quasi-static, since the evolution time of the characteristic structures exceeds the scale of their inhomogeneity divided by the velocity of typical particles, and the role of the induction electric field is inessential.

\section{2D simulations for a wide range of the plasma density and external magnetic field values. A comparative analysis}
\label{sec4}

\begin{figure}
	\centerline{\includegraphics[width=1.0\textwidth]{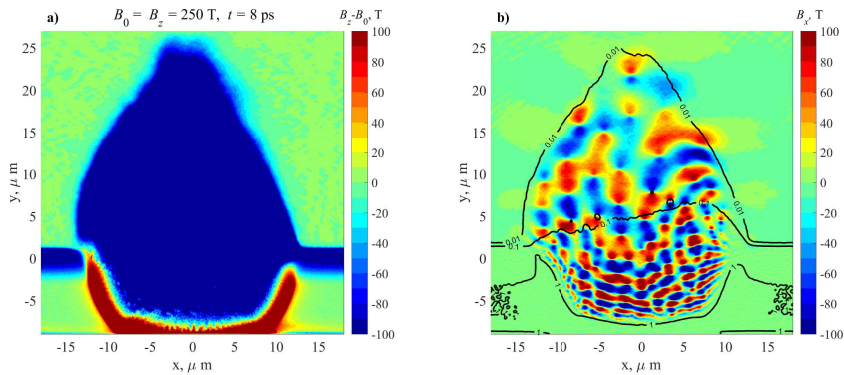}}
    \caption{
    Magnetic field structures generated in the 2D simulation of the expansion of the plasma cloud with hot electrons into vacuum at $t = 8$~ps $\approx 6 \cdot 10^4\omega_\mathrm{pe}^{-1}$ after the beginning of the expansion. The initial plasma density is $n_0 = 1.7 \cdot 10^{22}$~cm$^{-3}$. The external magnetic field is $B_0 = 250$~T and orthogonal to the simulation plane, $\vec{B}_0 \| Oz$. 
    (\textit{a})~$B_z$ component of the total magnetic field minus the external magnetic field. (\textit{b})~$B_x$ component of the magnetic field; black contours show the isolines of the normalized density for $n / n_0 = 0.01, 0.1, 1$.
    }
    \label{fig2}
\end{figure}

\begin{figure}
	\centerline{\includegraphics[width=1.0\textwidth]{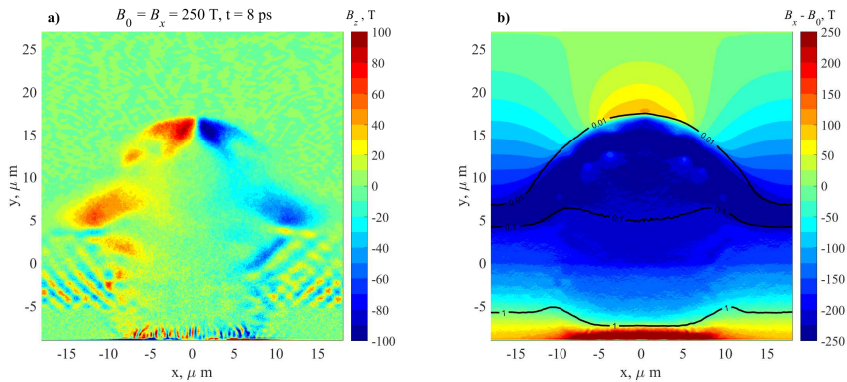}}
    \caption{
    Magnetic field structures generated in the 2D simulation of the expansion of the plasma cloud with hot electrons into vacuum at $t = 8$~ps after the beginning of the expansion. The initial plasma density is $n_0 = 1.7 \cdot 10^{22}$~cm$^{-3}$. The external magnetic field is $B_0 = 250$~T and lies in the simulation plane, $\vec{B}_0 \| Ox$.
    (\textit{a})~$B_z$ component of the magnetic field. (\textit{b})~$B_x$ component of the total magnetic field minus the external magnetic field; black contours show the isolines of the normalized density for $n / n_0 = 0.01, 0.1, 1$. 
    }
\label{fig3}
\end{figure}

\begin{figure}
	\centerline{\includegraphics[width=1.0\textwidth]{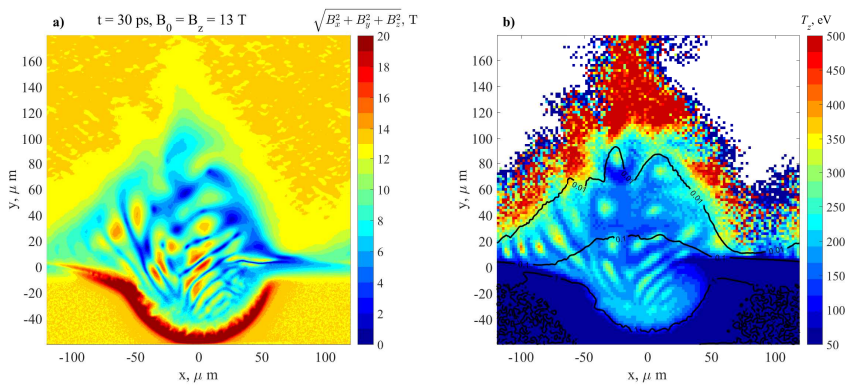}}
    \caption{
    The expansion of the plasma cloud with hot electrons (with the initial density of $n_0 = 10^{20}$~cm$^{-3}$) into vacuum with the external magnetic field $B_0 = 13$~T, $\vec{B}_0 \| Oz$, at $t = 30$~ps $\approx 2 \cdot 10^4\omega_\mathrm{pe}^{-1}$ after the beginning of the expansion.
    (\textit{a})~The magnitude of the total magnetic field. (\textit{b})~The effective temperature $T_z$ along the $z$-axis; black contours show the isolines of the normalized density for $n / n_0 = 0.01, 0.1, 1$.
    }
\label{fig4}
\end{figure}

\begin{figure}
	\centerline{\includegraphics[width=1.0\textwidth]{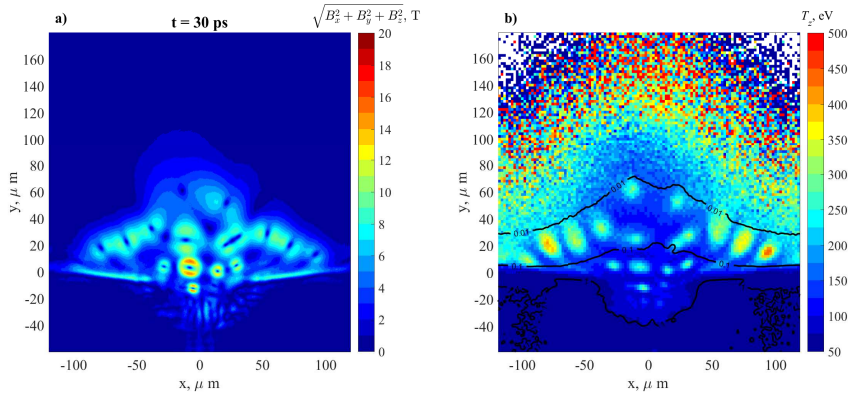}}
    \caption{
    The expansion of the plasma cloud with hot electrons (with the initial density of $n_0 = 10^{20}$~cm$^{-3}$) into vacuum without an external magnetic field at $t = 30$~ps after the beginning of the expansion.
    (\textit{a})~The magnitude of the total magnetic field. (\textit{b})~The effective temperature $T_z$ along the $z$-axis; black contours show the isolines of the normalized density for $n / n_0 = 0.01, 0.1, 1$.
    }
\label{fig5}
\end{figure}

\begin{figure}
	\centerline{\includegraphics[width=1.0\textwidth]{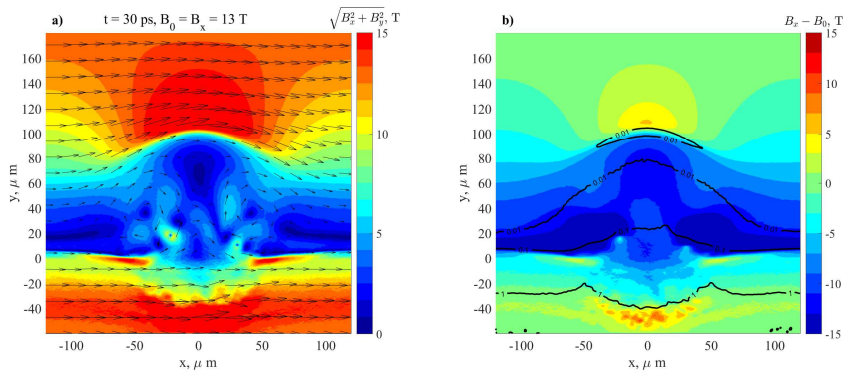}}
    \caption{
    The expansion of the plasma cloud with hot electrons (with the initial density of $n_0 = 10^{20}$~cm$^{-3}$) into vacuum with the external magnetic field $B_0 = 13$~T, $\vec{B}_0 \| Ox$, at $t = 30$~ps after the beginning of the expansion. 
    (\textit{a})~The magnitude of the total in-plane magnetic field; arrows show the direction of the in-plane field component. (\textit{b})~$B_x$ component of the total magnetic field minus the external magnetic field; black contours show the isolines of the normalized density for $n / n_0 = 0.01, 0.1, 1$.
    }
\label{fig6}
\end{figure}

The established validity of the simplified 2D3V calculations for a qualitative analysis of physical phenomena at a considerably long period of the decay of the plasma discontinuity with an elongated (semi-cylindrical) region of heated electrons, allows us to describe in more detail a number of features of this process at various plasma densities as well as various values and orientations of an external magnetic field. It is convenient to illustrate these features by comparing figs.~\ref{fig2}, \ref{fig3}, \ref{fig4}, \ref{fig5}, \ref{fig6}.

First of all, it is clear from figs.~\ref{fig1b}, \ref{fig2}, \ref{fig3}, and \ref{fig6} that, in the region of the expanding collisionless plasma cloud, the magnetic field could substantially weaken or significantly change its direction and even become directed opposite to the external one, keeping almost the same order of magnitude ($\sim B_{0}/3$). Moreover, under certain conditions, according to the laws of magnetostatics, in the outside boundary region next to the expanding cloud (practically in vacuum) the magnetic field can also change its direction and even increase, differently in different regions and depending on the strength of the local quasi-surface currents and the orientation of the external magnetic field. At the same time, for both orthogonal orientations of the external field ($B_{0z}$ and $B_{0x}$), there is an axial asymmetry, i.e., a difference in directions of the generated magnetic fields, especially near the initial plasma discontinuity to the left and to the right with respect to the center of the heated section. This asymmetry arises due to the vertical, parallel to the $y$-axis, component of the electric current created by the central ''fountain'' of the escaping electrons and the currents flowing in the dense plasma and compensating the charge of these escaped electrons, as can be seen from the magnetic field distributions in figs.~\ref{fig2}b and \ref{fig3}a (on the ''fountain'' mechanism in the absence of an external magnetic field, as in fig.~\ref{fig5}, see, e.g., \citet{Sakagami1979, Kolodner1979, Albertazzi2015}). 

\begin{figure}
	\centerline{\includegraphics[width=1.0\textwidth]{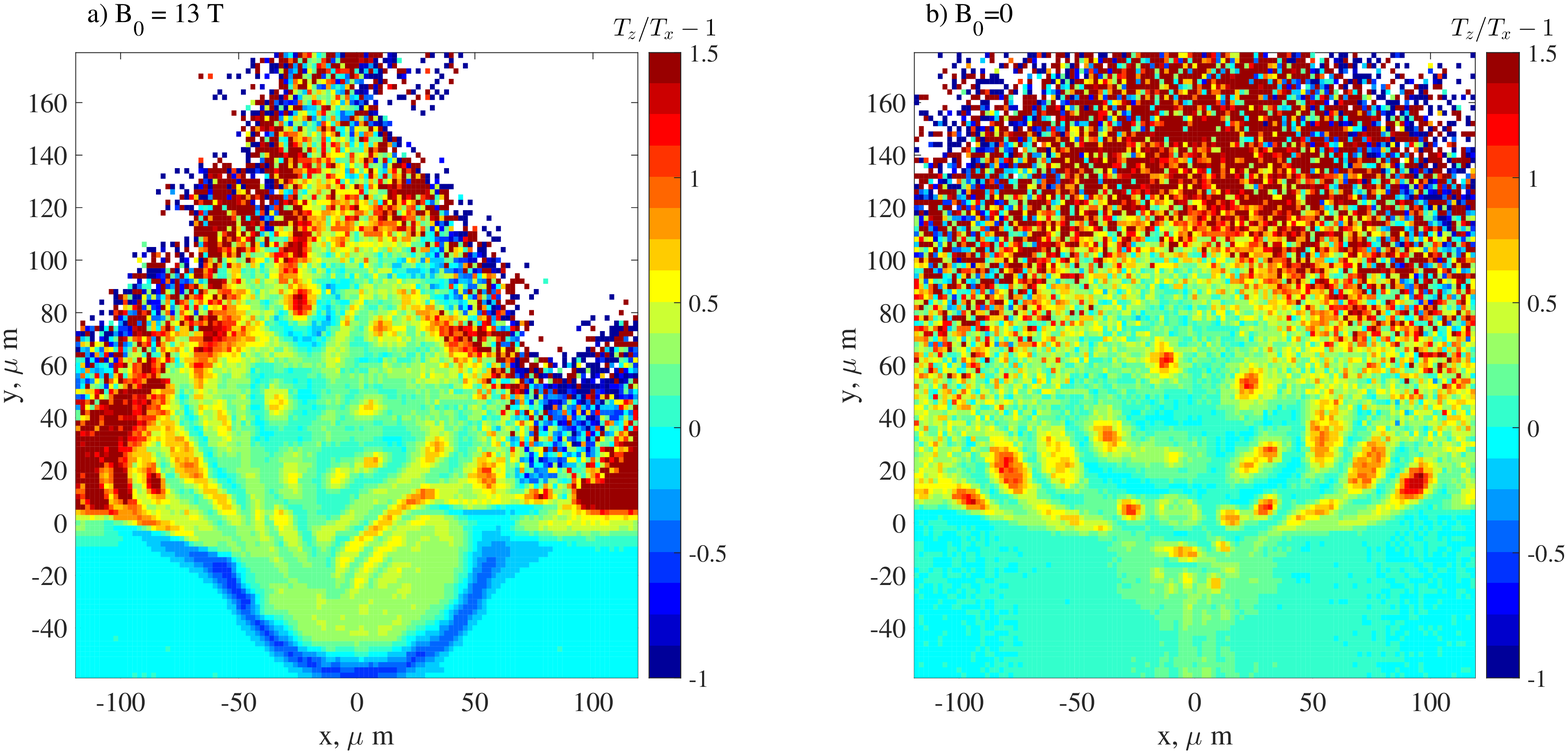}}
    \caption{
    The temperature anisotropy degree, $A = T_z/T_x-1$, induced by the expansion of the plasma cloud with hot electrons (with the initial density of $n_0 = 10^{20}$~cm$^{-3}$) into vacuum (\textit{a}) with the external magnetic field $B_0 = 13$~T, $\vec{B}_0 \| Oz$,
    (\textit{b}) without an external magnetic field, both at the moment of time $t = 30$~ps $\approx 2 \cdot 10^4\omega_\mathrm{pe}^{-1}$ after the beginning of the expansion.
    }
\label{fig7}
\end{figure}

According to the simulations, the resulting structure of the electric currents and quasi-magnetostatic fields is practically independent of an external magnetic field only if the latter is very weak ($B_0 < 0.1$~T and $B_0 < 1$~T for the chosen plasma parameters with the initial hot electron density of $n_0 = 10^{20}$~cm$^{-3}$ and $1.7 \cdot 10^{22}$~cm$^{-3}$, respectively). Such a weak field is easily displaced by the plasma and does not essemtially change an electron distribution function. If the external magnetic field is stronger, though still satisfies the inequality $B_0^2 / 8\pi \ll n_0 T$ and is displaced by the plasma, then the electron distribution function in a considerable volume of a plasma cloud is significantly changed during the plasma expansion. Nevertheless, the Weibel instability will proceed qualitatively similar to that in the absence of the external field, if both the electron gyrofrequency $e B_\mathrm{r} / (m c)$ for the remaining in the region (after the displacement) average magnetic field $B_\mathrm{r}$ and the inverse gyroradius of energetic electrons in this field are sufficiently (logarithmically) less than the maximum growth rate and the corresponding wavenumber of the instability in the anisotropic expanding plasma, respectively. Otherwise, the instability is influenced by the remaining large-scale inhomogeneous magnetic field which depends on the orientation of the external field and is comparable to the Weibel saturation field. As a result, the structure of the self-generated magnetic field during the decay of the discontinuity will differ considerably from that in the absence of an external field (fig.~\ref{fig5}).

The resulting profile of the ejected plasma cloud is almost independent of even stronger external fields for any of their orientations, e.g., for the values $B_{0z}$, $B_{0x}$ under $13$~T at the density of $n_0 = 1.7 \cdot 10^{22}$~cm$^{-3}$. However, upon reaching the indicated (see figs.~\ref{fig4} and \ref{fig6} for $n_0 = 10^{20}$~cm$^{-3}$) and larger values (for example, $B_{0z}$, $B_{0x} = 250$~T, as in figs.~\ref{fig2} and ~\ref{fig3} for $n_0 = 1.7 \cdot 10^{22}$~cm$^{-3}$), the cases of orientation of the external field along and across the semi-cylinder of heated electrons begin to differ significantly. In the first case, the aforementioned cumulative effect takes place and, as a result, the rate of the displacement of this field is significantly higher (for example, almost by $1.5$ times at $B_{0z} = 250$~T) and the plasma cloud is narrower (also almost by $1.5$ times at $B_{0z} = 250$~T) than in the case of the same-magnitude field $B_{0x}$ orthogonal to the heated semi-cylinder. In the latter case, the field $B_{0x}$ flattens the plasma cloud via a formation of a completely different system of the large-scale electric currents. In addition, only when the external field is oriented along the $z$-axis there is a noticeable violation of the symmetry of expansion caused by a systematic displacement of a large fraction of electrons to the left under the action of the Lorentz force (directed mainly opposite to the $x$-axis), see figs.~\ref{fig2} and \ref{fig4}. At the same time, only when the external field is oriented along the $x$-axis, a formation of the large-scale currents predominantly directed along the $z$-axis takes place at the top of the ejected plume. As a result, they stretch and smoothly bend the magnetic field lines in the $xy$-plane and significantly enhance the external magnetic field in vacuum above the plasma plume; see figs.~\ref{fig3}b and~\ref{fig6}b.

Our calculations reveal small-scale structures of electron currents flowing mainly parallel to the $z$-axis in the form of $z$-pinches with a wide range of transverse scales $\sim 2-50$~$\mu$m (deformed by the plasma density gradient). They gradually drift along with the plasma in the course of its expansion and give rise to an inhomogeneous set of dipole spots in the structure of the transverse magnetic field components $B_x$, $B_y$ (minus the external field $B_{0x}$, if any; see figs.~\ref{fig2}b and~\ref{fig3}b). 
The observed phenomenon is due to the Weibel instability owing to the anisotropically cooling hot electrons, whose effective temperature rapidly decreases along the $x$- and $y$-axes during the discontinuity decay and changes more slowly along the $z$-axis because of an unlimited length of the initially heated plasma semi-cylinder \citep[cf.][]{Thaury2010}. The typical patterns of the electron anisotropy in the cases with and without the external magnetic field $B_{0z}$ are shown in fig.~\ref{fig7}. The anisotropy degree is quite strong, $A \sim 1$, and the external field $B_{0z} = 13$~T does not inhibit the formation of a set of $z$-pinches, though introduces an asymmetry and slows down a bit the plasma expansion.
The appearance of the small-scale current filaments, similar to $z$-pinches, can be prevented by the strong external field $B_{0x}$ during the displacement of which the large-scale currents along the $z$-axis are generated in the plasma cloud and, as a result, the growth of the temperature anisotropy of the cooling electron distribution is suppressed. 

On the contrary, the presence of the external field $B_{0z}$ does not prevent the anisotropic cooling of electrons, so that, from the very beginning of the discontinuity decay, a formation of the multiple-pinch-like current structures occurs (see fig.~\ref{fig3}b for the $x$-component of the quasi-magnetostatic turbulent field created by them as well as fig.~\ref{fig4}b for the effective electron temperature $T_z$ along the $z$-axis associated with such a turbulence). The thickness of these $z$-pinches turns out to be of the order of the electron gyroradius, and their transverse magnetic fields can be of the order of or exceed the external field $B_{0z}$. In particular, in the case of fig.~\ref{fig4}, where $n_0 = 10^{20}$~cm$^{-3}$ and $B_{0z} = 13$~T, for a typical filament we observe the magnetic field magnitude $\sim 5-10$~T, radius $\sim 3-5$~$\mu$m and total current $\sim 100$~A which are an order of magnitude weaker, by two-three times larger and of the same order of magnitude than, respectively, those in the case of fig.~\ref{fig2} (or fig.~\ref{fig1b}), where both $n_0 = 1.7 \cdot 10^{22}$~cm$^{-3}$ and $B_{0z} = 250$~T are larger. The simulations for the intermediate value of the initial plasma density, $n_0 = 10^{21}$~cm$^{-3}$, show qualitatively the same results.

Additional simulations (not provided in this short article) show that an increase of the ion-to-electron mass ratio decreases the ion-acoustic speed, slows down the process of the plasma expansion and proportionally reduces the scale of the current structures being formed, but does not qualitatively change their small-to-large scale hierarchy and evolution. Also, as expected, an insufficient heating of electrons, i.e., the presence of a significant fraction of cold electrons, notably changes the density profile and geometry of the plasma ejection, the type of deformation and the rate of the displacement of the external field, the number and spatial distribution of the formed small-scale $z$-pinches.

\section{Main qualitative results and some open physical problems}
\label{sec5}

It is clear from the above that the process of the decay of a plasma--vacuum discontinuity with a semi-cylindrical region of heated electrons elongated along the plasma surface largely depends on the magnitude and direction of the external magnetic field parallel to the discontinuity surface, even if the pressure of this field is much less than the plasma pressure. The hot electrons make a decisive contribution to the latter and strongly affect the evolution and spatial structure of emerging quasi-magnetostatic perturbations of various scales, especially due to the anisotropic cooling during the expansion.

Along with the obvious influence of the sufficiently strong external magnetic field on the density profile of the expanding plasma cloud, it turns out that even a relatively weak external magnetic field can both suppress and promote, depending on its orientation, the formation of various current filaments, sheets and large-scale structures. The performed simulations and estimates allow us to determine the conditions under which the multiple formation, prolonged existence and significant shift of the localized current filaments (similar to $z$-pinches) and more complex current configurations take place in the most part of the region swept from the external magnetic field by the expanding plasma.

The resulting current filaments of the $z$-pinch type can have a significantly increased plasma density and create small-scale magnetic fields which turn out to be of the order of or even stronger than the external field and contain about several percent of the initial energy of hot electrons. This occurs in those spatio-temporal regions where the plasma density and the anisotropy of the electron velocity distribution are sufficiently high and promote the development of the Weibel-type instability until its nonlinear saturation as well as a long-term existence of the entire current structures.

The formation of magnetic fields with a larger scale, of the order of the characteristic transverse size of the elongated heated plasma region, is caused by both the initially strongest ''fountain'' currents of the fastest escaping electrons as well as by the rapidly generated volumetric or quasi-two-dimensional currents of the hot electrons moving inside the inhomogeneously expanding plasma cloud and along its boundary with the unperturbed external magnetic field. Interacting with this larger-scale current structure, the unidirectional external magnetic field oriented transverse to the main flow of the hot electrons can lead to not just a violation of the symmetry of the expansion, but also to a cumulation of the plasma flow and an inhomogeneous deformation of the emerging multi-scale current structures.

We considered just the simplest possible model of the expansion of an inhomogeneously heated magnetized plasma with a uniform density, bearing in mind an ideal cylindrical lens and an ideal target subjected to an ablation by a femtosecond laser beam. Obviously, rather different (not quasi-one-dimensional) distributions of the density and effective temperatures of the rapidly heated electrons are possible if they are exposed to ultrashort laser pulses of various duration, cross section, polarization, and optical frequency. Moreover, the velocity distribution function of electrons could be non-Maxwellian and anisotropic at the very beginning of the expansion. In the actual experiments, a non-planar initial geometry of the interface between a heated plasma and vacuum or a background plasma, particle collisions, and an inflow of hot electrons from deeper heated regions of the target~--- all of which we did not take into account~--- could also be important. These and other factors related to the planned experiments on the plasma discontinuity decay in an external magnetic field will be addressed elsewhere.

The obtained results can find applications in the topical studies of the astrophysical objects as well as the high-energy density laboratory plasma systems. The described transient phenomena of the self-consistent growth and nonlinear evolution of the small- and large-scale magnetic field structures are particularly relevant to the dynamics of the solar flares, stellar wind formation and nonstationary structure of planetary magnetosheath regions; see, e.g., \citep{DeForest2018, Dudik2017, Viall2020, Lazar2022, Voros2017, Kelley2009, Balogh2018}. Discussions of the related kinetic phenomena in the physics of cosmic plasma are beyond the scope of the present Letter.

\section{Conclusions}
\label{sec6}

The results presented above show that the character of the expansion of a collisionless electron-ion plasma into vacuum and the currents and magnetic fields of various scales generated in the transition layer depend significantly on the geometry of the region of initially heated electrons and the magnitude and direction of the external magnetic field oriented along the plasma surface. We carried out the detailed numerical (PIC) analysis for a region with initially isotropically heated electrons in the form of a long semi-cylinder, the axis of which is located on the plasma surface. This analysis discloses a crucial role of the Weibel-type instabilities that are associated with the emerging anisotropy of the electron velocity distribution and strongly depend on the orientation of the external magnetic field. 

We reveal, under certain conditions, a formation and a rapid expansion of highly inhomogeneous electron currents in the form of filaments (similar to $z$-pinches) parallel to the external magnetic field as well as a formation and a slow evolution of current sheets oriented at different angles to the boundary between the plasma and the deformed magnetic field. We find that these currents can create fields significantly exceeding in magnitude the external magnetic field and indicate qualitatively the conditions on the magnitude and orientation of the latter as well as on the plasma parameters and electron heating required for this to occur. The predicted phenomena of the decay of the plasma discontinuity are feasible not only in the laser plasma but also in the coronal arches, stellar wind, and explosive processes in the planetary magnetospheres. 

\section{Acknowledgments}

The laboratory-astrophysics part of the work, especially related to the 3D simulations, was supported by the Russian Science Foundation, project no. 21-12-00416. 
Simulations were carried out in the Joint Supercomputer Center of~the Russian Academy of~Sciences.

\bibliographystyle{jpp}
\bibliography{biblio}

\end{document}